# Controlled length-dependent interaction of Majorana modes in Yu-Shiba-Rusinov chains

**Lucas Schneider[1], Philip Beck[1], Jannis Neuhaus-Steinmetz[1], Thore Posske[2,3], Jens Wiebe[1,*] and Roland Wiesendanger[1]**

[1]Department of Physics, Universität Hamburg, D-20355 Hamburg, Germany.

[2]I. Institute for Theoretical Physics, Universität Hamburg, D-20355 Hamburg, Germany.

[3]The Hamburg Centre for Ultrafast Imaging, Luruper Chaussee 149, 22761 Hamburg, Germany.

*E-mail: jwiebe@physnet.uni-hamburg.de

**Magnetic atoms on superconductors induce localized Yu-Shiba-Rusinov (YSR) bound states. The proposal that topological superconductivity and Majorana modes can be engineered in arrays of hybridizing YSR states has led to their intense investigation. Here, we study Majorana modes emerging from bands of hybridized YSR states in artificially constructed Manganese (Mn) chains on superconducting Niobium (Nb). By controlling the chain geometry on the single atom level, we can measure the interaction-induced energy splitting of Majorana modes from both chain's ends with increasing chain length. We find periodic lengths where their interaction is tuned to zero within the experimental energy resolution. Our work unravels ways to manipulate and minimize interactions between Majorana modes in finite-size systems as required for Majorana-based storage and processing of quantum information.**

Realizing Majorana modes (MMs) as zero-energy excitations in solid state systems has been an immense quest in the past two decades, being motivated by their possible use for fault-tolerant topological quantum computing[1–3]. Theoretical proposals combine superconductivity, magnetism and Rashba spin-orbit coupling (SOC)[4–9]. Experimental platforms in one dimension featuring these effects include semiconducting nanowires in proximity to superconductors with an externally applied magnetic field[10,11] or atomic chains with ferromagnetic[12,13] or spin-helical order[14,15] on superconducting substrates. MMs on the system's boundaries are the consequence of a topologically non-trivial band structure in the chain's bulk. This makes them immune to perturbations sufficiently local as compared to the size of the system. So-far studied atomic chains are short, consisting of only tens of atoms[12,14,15]. Here, the MMs on two sides of the chain will inevitably interact, thereby splitting in energy away from zero. Indeed, Coulomb blockade spectroscopy has provided evidence for a small MM splitting depending on the device length in InSb nanowires[16]. However, experiments to measure this effect on magnetic atomic chains in real space turn out to be challenging due to a finite, mostly temperature limited energy resolution. Since MMs are expected to have a large spatial extent if SOC is weak[17], their hybridization should be stronger and might become experimentally accessible using comparably light superconducting substrates. In the following, we demonstrate that chains of Mn atoms on a clean Nb(110) surface exhibit an effect compatible with extended hybridizing MMs.

### Prediction of topological superconductivity in YSR chains

Topological superconductivity and the resulting Majorana modes can be engineered in one-dimensional ferromagnets with an odd number of spin-polarized bands crossing the Fermi energy $E_F$[4,5,8]. The low-energy bands may be formed by hybridizing Yu-Shiba-Rusinov (YSR)[6,8,9,18–21] states locally induced by magnetic impurities on superconducting substrates. This has led to intense investigations of YSR states in the past years[21–29]. In the following, we present the first experimental realization of a topological superconductor which can be conclusively derived from single, hybridizing YSR states in a bottom-up approach. Using scanning tunneling microscopy (STM) and spectroscopy (STS) with a superconducting tip to probe the local density of states (LDOS) at sub-gap energies (see Methods), it is shown in Fig. 1a that multiple YSR states are induced by single Mn atoms on clean Nb(110)[30]. Their spatial anisotropy facilitates different hybridization of the YSR states stemming from neighboring Mn atoms by tailoring the directionality of nanostructures on the Nb(110) surface, as it has been shown for dimers of Mn atoms[30]. Especially, the lowest energy YSR state – referred to as $\delta$ in the following – is extended along the [1$\bar{1}$0] direction (see Fig. 1a, right panel). Thus, it is promising to construct chains along the

[1$\bar{1}$0] direction (as sketched in Fig. 2a). We expect this orientation to lead to a dominant hybridization of the lowest energy $\delta$-YSR states compared to the weaker coupling of all other higher energy YSR states, especially because the inter-atomic distance in this configuration is comparably large ($d = 0.467$ nm). Similarly, Mn is in the center of the transition metal series and its $d$-states are energetically located at very high energies away from $E_F$. Even when hybridizing, the bandwidth of the emerging $d$-bands is expected to be too small to reach $E_F$. Thus, ideally, the low energy band structure would be reduced to an effective one-YSR-band system around $E_F$. In this case, sufficient hybridization between the $\delta$-states results in a topological band structure irrespective of the model parameters, strongly reminiscent of the seminal Kitaev chain model for topological superconductivity[1]. The magnetic moments in Mn chains along the [1$\bar{1}$0] direction are ferromagnetically aligned (see Supplementary Note 1), thus providing all necessary ingredients for topological superconductivity in the presence of any non-vanishing effective Rashba SOC $k_h$. With the use of model calculations approximating the effective low-energy theory of a one-dimensional chain of dilute YSR impurities[8,9] (see Methods and Supplementary Note 2 for details), we discuss the expected topological properties of chains crafted from single YSR atoms. Within this model, the chain is embedded in a three-dimensional superconductor. We additionally find very similar results using a tight-binding model for a chain on the surface (see Supplementary Note 3). We start by modelling the $\delta$-YSR states of a single Mn impurity[30] (see Methods and Fig. 1a) and use the same parameters to extrapolate to the case of a YSR chain. The resulting phase diagram shown in Fig. 1b demonstrates that the chain is indeed almost always in a topologically non-trivial phase[8]. This holds as long as the chain is sufficiently dilute to remain in an effective one-band scenario where only the hybridizing $\delta$-YSR states are relevant and as long as the effective coherence length $\xi$ of the substrate is not unrealistically small or its Fermi wavevector $k_{F,0}$ has a very specific size. Experimentally, we can determine $k_{F,0}$ to be $(0.6 \pm 0.1)\,\pi/d$ (see Supplementary Note 4), i.e. far from these critical points.

## Signatures of hybridizing MMs in Mn chains on Nb(110)

In order to experimentally realize this concept, Mn$_N$ chains consisting of $N$ atoms along the [1$\bar{1}$0] direction (Fig. 2a) were constructed by controlled lateral manipulation of Mn atoms on the Nb surface using the STM tip (see Methods). In Fig. 2b, we present an example of a Mn$_{32}$ chain and spatially resolved deconvoluted differential conductance (d$I$/d$V$) maps around it. We find states at zero energy that are well localized at the chain's ends with additional small LDOS oscillations in the interior of the chain. In contrast, energetically higher states (0.5 meV < $|E|$ < 1.5 meV) are distributed all over the chain. Spectra from the dataset in Fig. 2b measured at the chain's end and center as well as on the bare Nb substrate (Fig. 2c) reveal a narrow zero-energy peak in the d$I$/d$V$ signal localized on the chain's end, corresponding to the zero-energy state of Fig. 2b. Peaks corresponding to the finite energy states in Fig. 2b are distributed over the entire chain. The clearly resolved zero energy end state provides first evidence for MMs in the Mn chain.

Since we construct the chains atom-by-atom, we are able to track the changes in the low-energy states for each length $N$. As an example, we show the deconvoluted d$I$/d$V$ signal along the chain in a one-dimensional line of spectra, called d$I$/d$V$ line-profile in the following, for $N$ = 14 to 16 in Fig. 3a. Interestingly, we find similar zero-energy end states as in Fig. 2b for $N$ = 14 and $N$ = 16 (see arrows), separated from higher energy states by $\Delta' = 400$ μeV. Instead, for $N$ = 15, there are two states with a similarly strong localization on the chain's ends (see arrows) but split by $E_{hyb} \approx 300$ μeV symmetrically around $E_F$. We interpret these states as MMs localized on the two ends, but with a residual coupling due to the finite length of the chain, as will be substantiated below. If this is the case, the coupling will depend not only on the length of the chain, but also on the wavefunction modulation of the MMs. In order to investigate this effect, we show the deconvoluted d$I$/d$V$ signal measured at the end of another, structurally identical chain with varying chain length $N$ in Fig. 3b. With increasing $N$, we added Mn atoms to one chain end and measured d$I$/d$V$ spectra at the unperturbed end to trace the states' energies. In accordance with Fig. 3a, we find that the energy of the state closest to $E_F$, which corresponds to the end state, is modulated with a period of $\Delta N \approx 2$. The trend continues up to the longest chains we have built ($N$ = 45). This is most apparent when we plot the chains with even- and odd-$N$ separately (Fig. 3c). Here, the change in sub-gap state energies appears to be continuously changing as a function of $N$. Most notably, for certain chain lengths, such as $N$ = 12, 21, 32 and 42, the energy of the end state can be tuned to zero within the experimental peak width, which corresponds to $\Delta E = 50$ μeV.

## Theoretical modeling of finite size chains

To substantiate that the observed end states are indeed hybridizing MMs from the two ends of the chain, we performed the aforementioned model calculations[8,9] to simulate chains of $N$ sites in contact with a superconducting host. We find regimes of the model qualitatively reproducing the experimental data on finite chains well, as presented in Fig. 4. Using parameters yielding a band structure of the YSR chain as shown in Fig. 4a, we find end states at zero energy with a strong localization at the terminal sites for special lengths of the chain (Fig. 4b). Most notably, the $\Delta N \approx 2$ modulation of the low-energy states is in good agreement with the experiment (cf. Figs. 3b,c and Figs. 4c,d). The modulation turns out to be induced by a particular position of the Fermi points in the low-energy band structure: Since the YSR band crosses $E_F$ at $k_F \approx \pm \pi/2d$, the Fermi wavelength $\lambda_F = 2\pi/k_F \approx 4d$ is specifically related to the lattice constant. This leads to a modulation of eigenenergies in chains of length $N$, such as the magnetic chain, with $\Delta N \approx 2$. This type of beating effect in a quantum-size limited system has been observed on other platforms, e.g., quantum well states in thin films of Pb/Si(111)[31–33] or in predictions for Andreev-bound states in superconducting carbon nanotubes[34]. Equally, the effect can be understood in terms of interacting MMs: It has been shown that the emergent MMs feature a wavefunction modulation with $\lambda_F/2$ [35], as can be seen in Fig. 4b. Accordingly, the overlap and thus hybridization of MM wavefunctions from both ends of the chain are expected to oscillate with the total chain length with $\Delta N \approx 2$. Particularly, for chain lengths where the two wavefunctions are perfectly shifted by a π-phase difference, their overlap is tuned to zero and the MMs do not hybridize.

**Discussion**

The $\Delta N \approx 2$ modulation appears to originate from a specific sub-gap band structure with a steep band crossing $E_F$ at $k_F \approx \pm \pi/2d$. The agreement with our model calculations indicates that the relevant band is indeed formed by hybridizing $\delta$-YSR states of the single Mn atoms expanded along the $[1\bar{1}0]$ direction, which have energies close to zero for isolated atoms (Fig. 1a)[30]. Unlike the $\delta$-YSR states, sub-gap states appearing at higher energies $|E| > 1$ meV in Figs. 3b,c exhibit strong particle-hole asymmetric spectral weight and a different spatial distribution, indicating that they stem from other bands.

The question arises, whether topologically trivial explanations for the observed end states exist. The fact that both ends of the chain change equally when perturbing only one side (as shown in Fig. 3a and Supplementary Note 5) proves that the end state is a collective mode of the one-dimensional structure. We can therefore rule out that the end states are zero-dimensional features induced by local defects or localized YSR states. It is possible that the localization of the wavefunction closest to $E_F$ is less pronounced than the experiment suggests (see Supplementary Notes 2, 3 and 5). Especially since both YSR states and MMs can be significantly located in the superconducting host, the measurement of the LDOS above the atomic chain could suppress the signal in the chain's interior while amplifying the intensity at the chain's ends[13]. Note that the topological phase of the infinite system would still be non-trivial in this case. Topologically trivial phases could only be compatible with the experiment in the presence of additional low-energy bands. In this scenario, an even number of MMs from different bands would inevitably hybridize, thereby lifting their degeneracy and destroying topological protection. Experimentally, all features from additional bands are well separated from $E_F$ (see Figs. 3b,c), providing strong evidence that our chains indeed realize an effective one-band model in the low-energy limit. As such, our model calculations reveal that the system is topologically non-trivial in the relevant parameter regime (see Fig. 1b and Supplementary Note 2).

We emphasize, however, that the emergent MMs only experience a topological protection of the size of the bulk topological gap $\Delta_p$. The topological gap is expected to be significantly smaller than the observed MM energy splitting $E_{hyb}$ (see Supplementary Fig. 2e and Supplementary Note 2) and the finite size gap $\Delta'$. The long-range extension of the MMs has been shown to be inversely related to $\Delta_p$[17]. Our results indicate that the observation of a well localized zero-energy end state does not directly imply that the corresponding MMs remain non-interacting under the influence of small perturbations. Although the bulk topological phase of the system is non-trivial, there is no clear way to perform braiding experiments within the accessible chain lengths. We expect the energy of the MMs to converge to zero for very long chains with $N > 70$ only (see Supplementary Fig. 2).

**Outlook**

One way to improve the localization further and to reduce the hybridization of MMs would be to enhance the Rashba SOC in the system, either using a different superconducting host, heavy-material interlayers or artificial SOC[5,36,37], all of which are expected to enhance $\Delta_p$. However, as seen from the topological phase diagram in Fig. 1b using the parameters from Fig. 4 (black dashed lines), while the system is deep in the topological phase, it is near to a gap closing at $k_F \approx \pm \pi/2d$ (see Supplementary Note 2). Note that, uncommonly, this gap closing is between two topologically non-trivial regions[8,9]. Despite constant SOC, this gap closing adds a previously disregarded constraint for realizing MMs in a hard protection gap $\Delta_p$ in future experiments. Thanks to the atomic-scale control of nanostructure fabrication by single atom manipulation, we can envision studies of MM hybridization in artificially created networks of interacting chains[38,39]. An example for a junction of two topological Mn$_{12}$ chains with varying inter-chain distance is shown in Figs. 5a,b. We can control the number of unoccupied adsorption sites $N_\emptyset$ with the precision of a single atom and analyze the energy of the state closest to $E_F$ (Fig. 5c, left panel, see Supplementary Note 5 for the full dataset). Its energy modulates with $\Delta N_\emptyset \approx 2$ again, providing evidence that the two chains are coupled through long-range interactions of the YSR states with a strength oscillating with $\lambda_F/2$. This oscillatory change of the energy of the end state with inter-chain distance is qualitatively consistent with model calculations using the model and the parameters from Fig. 4 and removing $N_\emptyset$ lattice sites of the YSR chain (Fig. 5c, right panel). However, a quantitative description of these interactions will require more advanced modelling. We emphasize that the observed energy splitting is an interplay of inter- and intra-chain MM interactions, where the latter is discussed above for single chains. Similar networks of topological chains may ultimately be needed to explore Majorana-based quantum computation in future studies and are the crucial next step in Majorana nanowire experiments.

## Methods
### Experimental procedures
All experiments were performed in a home-built STM facility operated at a temperature of $T$ = 320 mK [40]. We used a Nb(110) single crystal as a substrate, cleaned by high temperature flashes to $T$ > 2700 K [41]. Subsequently, single Mn atoms were deposited onto the cold surface ($T$ < 7 K), resulting in a statistical distribution of adatoms. Superconducting tips were created by indenting electrochemically etched W tips into the substrate, thereby picking up a large cluster of superconducting Nb. STM images were measured maintaining a constant tunneling current $I$ while applying a constant bias voltage $V_{DC}$ across the tunneling junction. For the measurement of differential tunneling conductance (d$I$/d$V$) spectra, the tip was stabilized at bias voltage $V_{stab}$ and current $I_{stab}$. Subsequently, the feedback loop was opened and the bias voltage was swept from -4 mV to +4 mV. The d$I$/d$V$ signal was measured using a standard lock-in technique with a small modulation voltage $V_{mod}$ (RMS) of modulation frequency $f$ = 4.142 kHz added to $V_{DC}$. The d$I$/d$V$ line-profiles and maps were acquired recording multiple d$I$/d$V$ spectra along a line or grid, respectively. All datasets shown in the main manuscript were measured using $V_{stab}$ = $V_{DC}$ = -6 mV, $I_{stab}$ = 1 nA and $V_{mod}$ = 20 µV. Superconducting Nb tips have been chosen in order to increase the effective energy resolution. The measured differential tunneling conductance d$I$/d$V$ is thus proportional to the convolution of the local density of states (LDOS) of the sample and the superconducting tip density of states (DOS). We show numerically deconvoluted STS data throughout the manuscript, resembling the sample's LDOS (see Supplementary Note 6 for details). The chains were assembled using lateral atom manipulation[42] techniques at low tunneling resistances of $R \approx$ 30 - 60 kΩ.

### Model for single and hybridizing YSR states
In the model of single YSR impurities embedded in a superconducting host (see Refs. 8 and 43), the sub-gap states are characterized by a magnetic scattering term $J$ and an additional non-magnetic scattering term $V$. Their energy $E(A,B)$ and particle-weight $P(A,B)$ can be written in terms of the dimensionless parameters $A = \pi \nu_0 J$ and $B = \pi \nu_0 V$, with the normal-phase density of states $\nu_0$ and the superconducting s-wave pairing $\Delta_s$:

$$E(A,B) = \Delta_s \frac{1 - A^2 + B^2}{\sqrt{(1 - A^2 + B^2)^2 + 4A^2}} \quad (1)$$

$$P(A,B) = \frac{1 + (A+B)^2}{1 + (A+B)^2 + 1 + (A-B)^2} \quad (2)$$

We find that the $\delta$-YSR states of Mn atoms on Nb(110)[30] are well reproduced by choosing $A = 1.1$ and $B = 0.2$ leading to the correct energy and particle-hole asymmetry of the experimentally measured peaks in d$I$/d$V$ (see fit with two Gaussians in Fig. 1a of the main text). Accordingly, these parameters are used for the description of chains. Note that the choice of $A = 0.94$ and $B = -0.2$ also reproduces the YSR peaks well and leads to very similar topological phase diagrams.

To describe chains of weakly interacting YSR atoms, we use a model based on a work of Pientka et al. [8] which is extended to include non-magnetic scattering at the YSR impurity (i.e. the $B$ term in equations (1) and (2)). The Bogoliubov-de-Gennes (BdG) Hamiltonian of a YSR chain with $n$ sites that are out-of-plane spin-polarized is

$$\mathcal{H} = \epsilon_p \tau^z + \Delta_s \tau^x + \sum_{j=1}^{n} (V\tau^z - J\sigma^z) \delta(\mathbf{r} - j\, d\, (1,0,0)^\mathrm{T}), \quad (3)$$

with the distance $d$ between the atoms, $\tau^i = \mathbf{1}_{2\mathrm{x}2} \otimes s^i$, $\sigma^i = s^i \otimes \mathbf{1}_{2\mathrm{x}2}$, the Pauli matrices $s^i$, the Kronecker product $\otimes$, the dispersion of the superconductor $\epsilon_p = (p^2/(2m) - \mu) + i\lambda(p_x \sigma^y - p_y \sigma^x)$ with the effective electron mass $m$, the chemical potential $\mu$ and momentum $p$ as well as Rashba SOC[43] of magnitude $\lambda$. $J$ and $V$ are defined above for the single YSR states and the BdG Hamiltonian $\mathcal{H}$ is acting on Nambu space in the basis $(c_\uparrow^\dagger(\mathbf{r}), c_\downarrow^\dagger(\mathbf{r}), c_\downarrow(\mathbf{r}), -c_\uparrow(\mathbf{r}))^\mathrm{T}$. Ref. 8 considered the case $V = 0$ and a helical magnetic texture. Our extended model recovers these results when noting that a magnetic helix described by the pitch $\pi/k_\mathrm{h}$ and an angle $\phi$ between neighboring spins is equivalent to an effective Rashba SOC parameter $\lambda = \hbar\phi/(2dm) = \hbar k_\mathrm{h}/m$ [36]. The presence of Rashba SOC in the experimental system can be motivated by the inevitably broken inversion-symmetry at a surface, where the chains in the experiment are located (see also Supplementary Note 3). We emphasize that the magnitude of SOC is not expected to be exactly zero, although the materials used (Nb and Mn) are comparably light. Following the lines of Ref. 8, an effective tight-binding model can subsequently be derived for the YSR bands in the low-energy limit. We refer to the original publication by Pientka *et al.* [8] for details of this derivation.

We derive a low-energy BdG Hamiltonian with on-site terms $h_{i,i}$, inter-site hopping terms $h_{i,j}$ ($i \neq j$), and effective p-wave superconductivity $\Delta_{i,j}$ ($i \neq j$) that read

$$h_{i,i} = \Delta_s \frac{\left(A - \sqrt{(A^2 - B^2)^2 + B^2}\right)}{(A-B)(A+B)}, \quad (4)$$

$$h_{i,j} = -\frac{\Delta_s\, e^{-\frac{d|i-j|}{\xi}} \cos[k_\mathrm{h}\, d\, (i-j)](m_{1,1}(A,B)\cos[k_{\mathrm{F},0}\, d\, |i-j|] + m_{1,2}(A,B)\sin[k_{\mathrm{F},0}\, d\, |i-j|])}{k_{\mathrm{F},0}\, d\, |i-j|}, \quad (5)$$

$$\Delta_{i,j} = -\frac{\Delta_s\, e^{-\frac{d|i-j|}{\xi}} \sin[k_\mathrm{h}\, d\, (i-j)](m_{2,1}(A,B)\cos[k_{\mathrm{F},0}\, d\, |i-j|] + m_{2,2}(A,B)\sin[k_{\mathrm{F},0}\, d\, |i-j|])}{k_{\mathrm{F},0}\, d\, |i-j|}. \quad (6)$$

Here, $k_{\mathrm{F},0}$ is the Fermi wavevector of the superconducting host in the metallic state, $\xi$ is the effective coherence length in the YSR chain and the coefficients $m_{1,1}$, $m_{1,2}$, $m_{2,1}$, and $m_{2,2}$ are

$$m_{1,1}(A,B) = \frac{B\left(2A(A-B)^2 + B - \sqrt{B^2 + (A^2 - B^2)^2}\right)}{A^4 - 2A^3B + 2A^2B^2 - 2AB^3 + B\left(B + B^3 - \sqrt{B^2 + (A^2 - B^2)^2}\right)} \quad (7)$$

$$m_{1,2}(A,B) = \frac{(A-B)(A^2 + B^2)(-B + \sqrt{B^2 + (A^2 - B^2)^2})}{(A+B)(A^4 - 2A^3B + 2A^2B^2 - 2AB^3 + B(B + B^3 - \sqrt{B^2 + (A^2 - B^2)^2}))} \quad (8)$$

$$m_{2,1}(A,B) = \frac{A^4 - B^4}{\sqrt{A^8 + 6A^2B^4 + B^6 + B^8 + A^4(B^2 - 2B^4) - 4A^3B^2\sqrt{B^2 + (A^2 - B^2)^2} - 4AB^4\sqrt{B^2 + (A^2 - B^2)^2}}} \quad (9)$$

$$m_{2,2}(A,B) = \frac{B\left(A^2 + B^2 - 2A\sqrt{B^2 + (A^2 - B^2)^2}\right)}{\sqrt{A^8 + 6A^2B^4 + B^6 + B^8 + A^4(B^2 - 2B^4) - 4A^3B^2\sqrt{B^2 + (A^2 - B^2)^2} - 4AB^4\sqrt{B^2 + (A^2 - B^2)^2}}} \quad (10)$$

This effective Hamiltonian acts on the basis of YSR states of the individual impurities featuring a particle-weight $P(A,B)$ (see equation (2))[44].

We compute the LDOS as a function of energy $E$ and position $x$ along a one-dimensional lattice of $N$ sites in Fig. 4 of the main text by diagonalizing the Hamiltonian in equations (4)-(6) and summing over all pairs of eigenvalues $E_i$ and eigenvectors $\psi_i$:

$$\text{LDOS}(E,x) = \sum_i \left[ P(A,B) |\psi_{i,e}(x)|^2 + (1 - P(A,B))|\psi_{i,h}(x)|^2 \right] \left( -\frac{\partial f(E - E_i, T = 320 \text{ mK})}{\partial E} \right) \quad (11)$$

with the respective particle- (e) and hole-components (h) of the solutions and the Fermi-Dirac function $f(E,T)$ simulating the experimental thermal broadening. To accurately obtain the particle-hole asymmetry of all states in terms of the physically original quasiparticles, which is measured in the experiment, $P(A = 1.1, B = 0.2)$ is multiplied with the particle-component of a state and $(1 - P(A = 1.1, B = 0.2))$ is multiplied with the hole-component. We obtain the band structure for an infinite chain by Fourier transforming the Hamiltonian with periodic boundary conditions applied[8]. For the numerical calculations in Fig. 4 of the main text, we used the parameters $A = 1.1$, $B = 0.2$, $k_h = 0.05\ \pi/d$, $k_{F,0} = 0.53\ \pi/d$, $\xi = 4.67$ nm, $d = 0.467$ nm, $\Delta_s = 1.5$ meV. The value for $k_{F,0}$ is - within the error bar - compatible with the value determined experimentally in Supplementary Note 4. Note that the only free parameters that cannot be directly determined experimentally from single YSR states are $k_h$ and $\xi$ (see Supplementary Note 2).

The topological invariant M is calculated via

$$\mathcal{M} = \text{sgn}\{\text{Pf}[\tilde{H}(0)]\text{Pf}[\tilde{H}(\pi)]\} \quad (12)$$

where Pf denotes the Pfaffian and $\tilde{H}(k)$ is the $k$-space Hamiltonian in the Majorana basis[1].

## Acknowledgements
We thank Stephan Rachel, Dirk Morr, Levente Rózsa, Elena Vedmedenko and Falko Pientka for helpful discussions. L.S., T.P., J.W., and R.W. gratefully acknowledge funding by the Cluster of Excellence 'Advanced Imaging of Matter' (EXC 2056 - project ID 390715994) of the Deutsche Forschungsgemeinschaft (DFG). J.N.-S. and R.W. gratefully acknowledge financial support from the European Union via the ERC Advanced Grant ADMIRE (project No. 786020). P.B., J.W. and R.W. acknowledge support by the Deutsche Forschungsgemeinschaft (DFG, German Research Foundation) – SFB-925 – project 170620586.

## Data availability
The authors declare that the data supporting the findings of this study are available within the paper and its supplementary information files.

## Competing interests
The authors declare no competing interests.

## Author contributions
L.S., P.B., J.W. and R.W. conceived the experiments. L.S. and P.B. performed the measurements and analyzed the experimental data. L.S., J.N.-S., and T.P. performed the model simulations. L.S. prepared the figures, L.S. and J.W. wrote the paper. All authors contributed to the discussions and to correcting the manuscript.

## References

1. Kitaev, A. Y. Unpaired Majorana fermions in quantum wires. *Physics-Uspekhi* **44**, 131–136 (2001).
2. Kitaev, A. Y. Fault-tolerant quantum computation by anyons. *Ann. Phys. (N. Y.)* **303**, 2–30 (2003).
3. Stern, A. Non-Abelian states of matter. *Nature* **464**, 187–193 (2010).
4. Li, J. *et al.* Topological superconductivity induced by ferromagnetic metal chains. *Phys. Rev. B* **90**, 235433 (2014).

5. Nadj-Perge, S., Drozdov, I. K., Bernevig, B. A. & Yazdani, A. Proposal for realizing Majorana fermions in chains of magnetic atoms on a superconductor. *Phys. Rev. B* **88**, 020407 (2013).
6. Schecter, M., Flensberg, K., Christensen, M. H., Andersen, B. M. & Paaske, J. Self-organized topological superconductivity in a Yu-Shiba-Rusinov chain. *Phys. Rev. B* **93**, 140503 (2016).
7. Klinovaja, J., Stano, P., Yazdani, A. & Loss, D. Topological Superconductivity and Majorana Fermions in RKKY Systems. *Phys. Rev. Lett.* **111**, 186805 (2013).
8. Pientka, F., Glazman, L. I. & von Oppen, F. Topological superconducting phase in helical Shiba chains. *Phys. Rev. B* **88**, 155420 (2013).
9. Pientka, F., Peng, Y., Glazman, L. & Oppen, F. Von. Topological superconducting phase and Majorana bound states in Shiba chains. *Phys. Scr.* **T164**, 014008 (2015).
10. Mourik, V. *et al.* Signatures of Majorana Fermions in Hybrid Superconductor-Semiconductor Nanowire Devices. *Science* **336**, 1003–1007 (2012).
11. Das, A. *et al.* Zero-bias peaks and splitting in an Al-InAs nanowire topological superconductor as a signature of Majorana fermions. *Nat. Phys.* **8**, 887–895 (2012).
12. Nadj-Perge, S. *et al.* Observation of Majorana fermions in ferromagnetic atomic chains on a superconductor. *Science* **346**, 602–607 (2014).
13. Feldman, B. E. *et al.* High-resolution studies of the Majorana atomic chain platform. *Nat. Phys.* **13**, 286–291 (2017).
14. Kim, H. *et al.* Toward tailoring Majorana bound states in artificially constructed magnetic atom chains on elemental superconductors. *Sci. Adv.* **4**, eaar5251 (2018).
15. Schneider, L. *et al.* Controlling in-gap end states by linking nonmagnetic atoms and artificially-constructed spin chains on superconductors. *Nat. Commun.* **11**, 4707 (2020).
16. Albrecht, S. M. *et al.* Exponential protection of zero modes in Majorana islands. *Nature* **531**, 206–209 (2016).
17. Peng, Y., Pientka, F., Glazman, L. I. & von Oppen, F. Strong Localization of Majorana End States in Chains of Magnetic Adatoms. *Phys. Rev. Lett.* **114**, 106801 (2015).
18. Yu, L. Bound state in Superconductors with paramagnetic impurities. *Acta Phys. Sin.* **21**, (1965).
19. Shiba, H. Classical Spins in Superconductors. *Prog. Theor. Phys.* **40**, (1968).
20. Rusinov, A. I. Superconductivity near a Paramagnetic Impurity. *ZhETF Pisma Redaktsiiu* **9**, 146 (1968).
21. Heinrich, B. W., Pascual, J. I. & Franke, K. J. Single magnetic adsorbates on s-wave superconductors. *Prog. Surf. Sci.* **93**, 1–19 (2018).
22. Ji, S.-H. *et al.* High-Resolution Scanning Tunneling Spectroscopy of Magnetic Impurity Induced Bound States in the Superconducting Gap of Pb Thin Films. *Phys. Rev. Lett.* **100**, 226801 (2008).
23. Yazdani, A., Eigler, D. M., Lutz, C. P., Jones, B. A. & Crommie, M. F. Probing the Local Effects of Magnetic Impurities on Superconductivity. *Science* **275**, 1767–1770 (1997).
24. Franke, K. J., Schulze, G. & Pascual, J. I. Competition of Superconducting Phenomena and Kondo Screening at the Nanoscale. *Science* **332**, 940–944 (2011).
25. Cornils, L. *et al.* Spin-Resolved Spectroscopy of the Yu-Shiba-Rusinov States of Individual Atoms. *Phys. Rev. Lett.* **119**, 197002 (2017).
26. Choi, D.-J. *et al.* Mapping the orbital structure of impurity bound states in a superconductor. *Nat. Commun.* **8**, 15175 (2017).
27. Ruby, M., Peng, Y., von Oppen, F., Heinrich, B. W. & Franke, K. J. Orbital Picture of Yu-Shiba-Rusinov Multiplets. *Phys. Rev. Lett.* **117**, 186801 (2016).
28. Schneider, L. *et al.* Magnetism and in-gap states of 3d transition metal atoms on superconducting Re. *npj Quantum Mater.* **4**, 42 (2019).
29. Odobesko, A. *et al.* Observation of tunable single-atom Yu-Shiba-Rusinov states. *Phys. Rev. B* **102**, 174504 (2020).
30. Beck, P. *et al.* Spin-orbit coupling induced splitting of Yu-Shiba-Rusinov states in antiferromagnetic dimers. *Nat. Commun.* **12**, 2040 (2021).
31. Qi, Y. *et al.* Atomic-layer-resolved local work functions of Pb thin films and their dependence on quantum well states. *Appl. Phys. Lett.* **90**, 013109 (2007).
32. Guo, Y. *et al.* Superconductivity modulated by quantum size effects. *Science* **306**, 1915–1917 (2004).
33. Upton, M. H., Wei, C. M., Chou, M. Y., Miller, T. & Chiang, T.-C. Thermal Stability and Electronic Structure of Atomically Uniform Pb Films on Si(111). *Phys. Rev. Lett.* **93**, 026802 (2004).
34. Crépin, F., Hettmansperger, H., Recher, P. & Trauzettel, B. Even-odd effects in NSN scattering problems: Application to graphene nanoribbons. *Phys. Rev. B* **87**, 195440 (2013).
35. Klinovaja, J. & Loss, D. Composite Majorana fermion wave functions in nanowires. *Phys. Rev. B* **86**, 085408 (2012).

36. Kjaergaard, M., Wölms, K. & Flensberg, K. Majorana fermions in superconducting nanowires without spin-orbit coupling. *Phys. Rev. B* **85**, 020503 (2012).
37. Desjardins, M. M. *et al.* Synthetic spin–orbit interaction for Majorana devices. *Nat. Mater.* **18**, 1060–1064 (2019).
38. Björnson, K. & Black-Schaffer, A. M. Majorana fermions at odd junctions in a wire network of ferromagnetic impurities. *Phys. Rev. B* **94**, 100501 (2016).
39. Alicea, J., Oreg, Y., Refael, G., von Oppen, F. & Fisher, M. P. A. Non-Abelian statistics and topological quantum information processing in 1D wire networks. *Nat. Phys.* **7**, 412–417 (2011).
40. Wiebe, J. *et al.* A 300 mK ultra-high vacuum scanning tunneling microscope for spin-resolved spectroscopy at high energy resolution. *Rev. Sci. Instrum.* **75**, 4871–4879 (2004).
41. Odobesko, A. B. *et al.* Preparation and electronic properties of clean superconducting Nb(110) surfaces. *Phys. Rev. B* **99**, 115437 (2019).
42. Schneider, L., Beck, P., Wiebe, J. & Wiesendanger, R. Atomic-scale spin-polarization maps using functionalized superconducting probes. *Sci. Adv.* **7**, eabd7302 (2021).
43. Bychkov, Y. A. & Rashba, E. I. Oscillatory effects and the magnetic susceptibility of carriers in inversion layers. *J. Phys. C Solid State Phys.* **17**, 6039–6045 (1984).
44. von Oppen, F., Peng, Y. & Pientka, F. *Topological superconducting phases in one dimension: Lecture Notes of the Les Houches Summer School*. *Topological Aspects of Condensed Matter Physics* (Oxford Univ. Press, 2014).

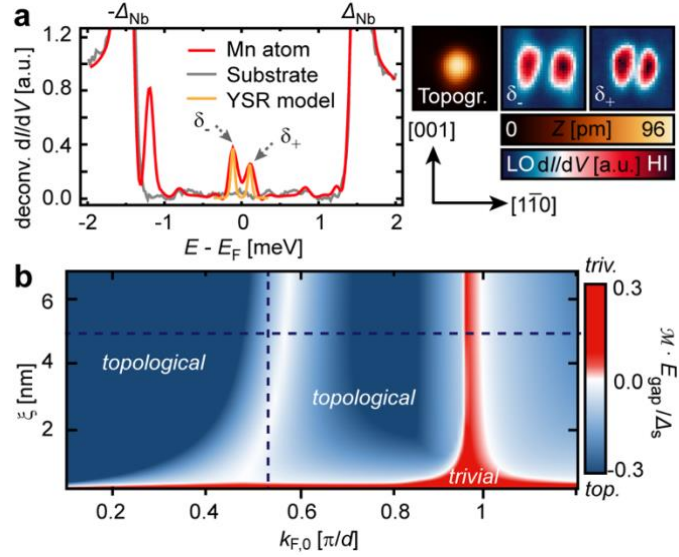

**Figure 1 | YSR states of single Mn atoms on Nb(110) and resulting topological properties of a hybridized chain. a**, Left panel: Deconvoluted d$I$/d$V$ spectra averaged around a single Mn atom on Nb(110) and on the bare substrate. The pair of YSR states $\delta_{+/-}$ closest to $E_F$ inside the energy gap of the superconducting substrate is highlighted by arrows. The measured coherence peaks are energetically located at the superconducting gap of bulk Nb $\Delta_{Nb}$ = 1.52 meV. Both the energy and the particle-hole asymmetry of the YSR states are well described by a classical YSR model with the magnetic and non-magnetic scattering parameters $A = 1.1$ and $B = 0.2$ (orange line, see Methods). Right panels: STM image of a single Mn atom (Topogr.) and the spatially resolved d$I$/d$V$ map around the atom at the energies of $\delta_{+/-}$, showing the state's lobes along [1$\bar{1}$0] which is the direction of the chains (image sizes are 2 x 2 nm$^2$). **b**, Topological phase diagram calculated for an infinite chain of interacting YSR states for different values of the Fermi vector $k_{F,0}$ and coherence length $\xi$ of the substrate. The parameters for the YSR states as derived in **a** are used, and a constant, weak effective Rashba SOC $k_h = 0.05\ \pi/d$ is added. The phase diagram does not change qualitatively for other small values of $k_h$. The color scale indicates the magnitude of the system's gap $E_{gap}$ multiplied by the topological $\mathbb{Z}_2$ invariant $\mathcal{M}$, which is $\mathcal{M}$ = -1 (topological, blue) or $\mathcal{M}$ = +1 (trivial, red), compared with the superconducting pairing of the substrate $\Delta_s$. The dark dashed lines mark the parameters used in Fig. 4 that describe the experimental data in Fig. 3 best.

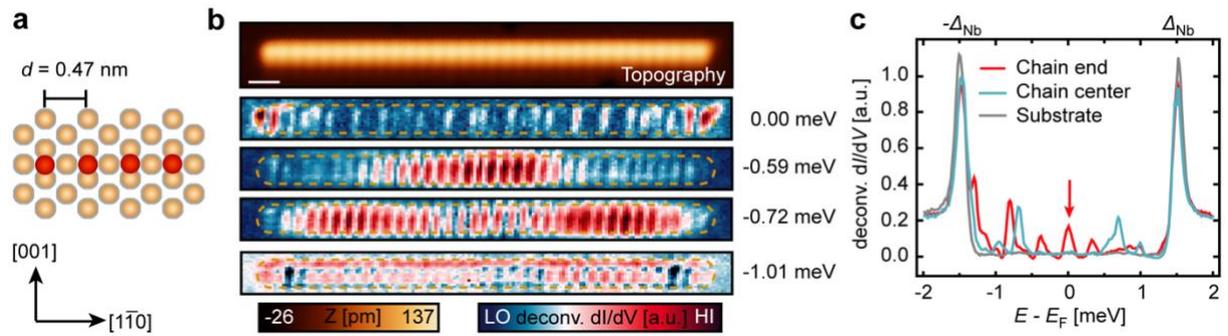

**Figure 2 | In-gap states in Mn chains on Nb(110) along [1$\bar{1}$0]. a**, Geometry of the experimentally assembled Mn atoms (red spheres) on top of the atoms of the superconducting Nb host (brown spheres). **b**, Constant-current STM image (Topography, top panel) of a Mn$_{32}$ chain and corresponding deconvoluted d$I$/d$V$ maps (bottom panels) at selected energies as indicated. The brown dashed lines in the d$I$/d$V$ maps mark the position of the chain. The white scale bar corresponds to 1 nm. **c**, Single deconvoluted d$I$/d$V$ spectra obtained on the chain's end, in the center and on the Nb substrate, respectively. The zero-energy peak is highlighted by the red arrow.

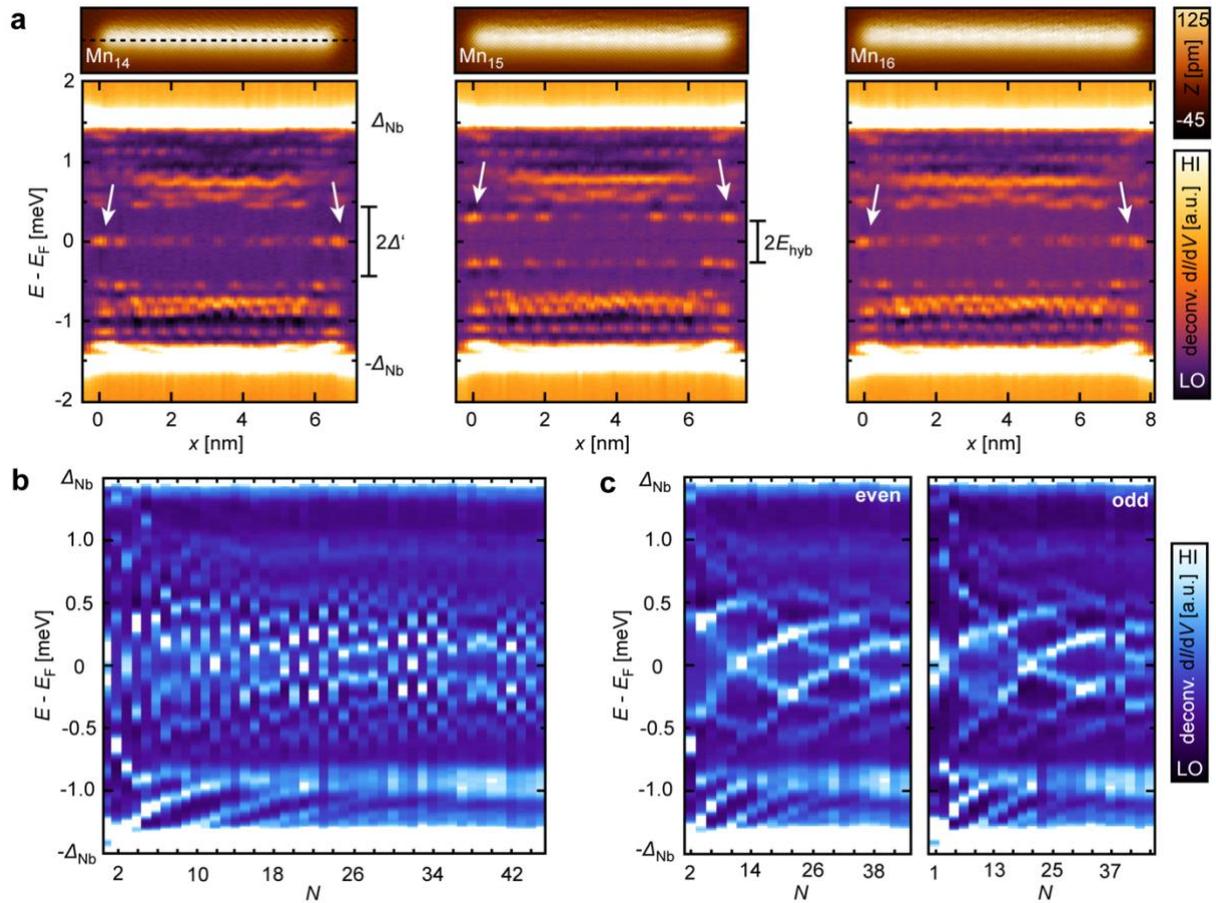

**Figure 3 | Chain-length dependence of in-gap states. a**, Top panels: STM images of Mn$_N$ chains with $N$ = 14, 15 and 16. Bottom panels: deconvoluted d$I$/d$V$ line-profiles acquired along the longitudinal axis through the center of the three chains (as indicated by the dashed line along the Mn$_{14}$ chain). The STM images are aligned with the d$I$/d$V$ line-profiles and edge states are highlighted by white arrows. **b**, Sequence of d$I$/d$V$ spectra obtained on the end of Mn$_N$ chains of different number of sites $N$ located at a different sample position as the chains in **a**. **c**, Dataset from **b** with the even- and odd-numbered chains plotted in separate panels.

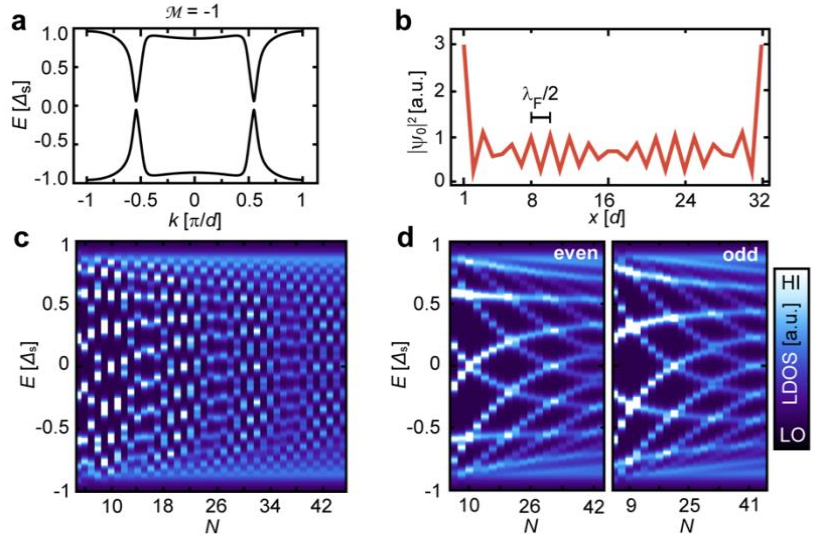

**Figure 4 | Theoretical model for interacting MMs derived from YSR bands. a**, Band structure of an infinite chain of YSR impurities using parameters describing the experimentally measured energy and particle-hole asymmetry of the $\delta$-YSR states. The topological index of the infinite chain is non-trivial ($\mathcal{M}$ = -1). **b**, Calculated zero-energy LDOS along a finite chain of 32 sites using the parameters from **a**. **c**, LDOS on the first site of a finite chain of length $N$. **d**, Dataset from **c** with the even- and odd-numbered chains plotted in separate panels. Parameters: $A = 1.1$, $B = 0.2$, $k_\mathrm{h} = 0.05\ \pi/d$, $k_{\mathrm{F},0} = 0.53\ \pi/d$, $\xi = 4.67$ nm, $d = 0.467$ nm, $\Delta_\mathrm{s} = \Delta_\mathrm{Nb} = 1.5$ meV, see Methods and Supplementary Note 2 for details.

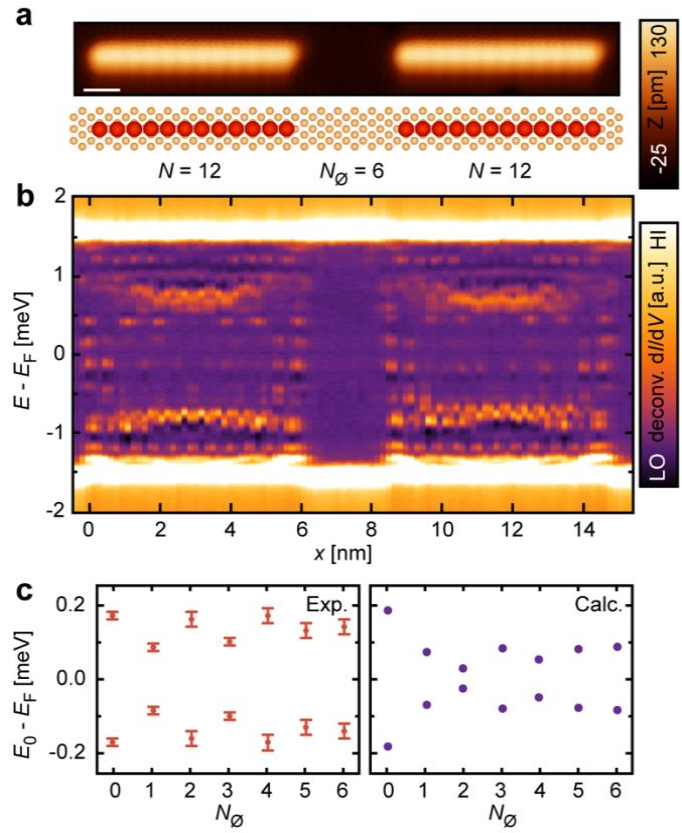

**Figure 5 | Junctions of interacting Mn$_{12}$ chains. a**, STM image (top) of two Mn$_{12}$ chains with $N_\emptyset = 6$ empty sites between the chains and a respective sketch of the geometric structure (bottom). The white bar corresponds to 1 nm. **b**, d$I$/d$V$ line-profiles along the longitudinal axes through the centers of both chains in **a** aligned with the topography. **c**, Experimentally measured (left) and calculated (right) energy $E_0$ of the lowest energy state vs. the inter-chain spacing $N_\emptyset$. The same model parameters as in Fig. 4 are used.